\documentclass[reprint,prl,preprintnumbers,amsmath,amssymb]{revtex4-1}
\usepackage{hyperref}
\usepackage{natbib}
\usepackage[dvips]{graphicx}
\usepackage{dcolumn}
\usepackage{bm}

\begin{document}
\bibliographystyle{unsrtnat}
\title{\large {Measurement of intrinsic Dirac fermion cooling on the surface of a topological insulator Bi$_2$Se$_3$ using time- and angle-resolved
photoemission spectroscopy}}

\vspace{0.5cm}
\author{Y. H. Wang$^{1,2}$}
\author{D. Hsieh$^{1}$}
\author{E. J. Sie$^{1}$}
\author{H. Steinberg$^{1}$}
\author{D. R. Gardner$^{1}$}
\author{Y. S. Lee$^{1}$}
\author{P. Jarillo-Herrero$^{1}$}
\author{N. Gedik$^{1}$}
\affiliation{$^1$Department of Physics, Massachusetts Institute of
Technology, Cambridge MA 02139, USA} \affiliation{$^2$Department of
Physics, Harvard University, Cambridge MA 02138, USA}

\vspace{0.5cm}

\begin{abstract}
We perform time- and angle-resolved photoemission spectroscopy
of a prototypical topological insulator Bi$_2$Se$_3$ to study the ultrafast dynamics of surface and bulk electrons after photo-excitation. By analyzing the evolution of surface states and bulk band spectra, we obtain their electronic temperature and chemical potential relaxation dynamics separately. These dynamics reveal strong phonon-assisted surface-bulk coupling at high lattice temperature and total suppression of inelastic scattering between the surface and the bulk at low lattice temperature. In this low temperature regime, the unique cooling of Dirac fermions in TI by acoustic phonons is manifested through a power law dependence of the surface temperature decay rate on carrier density.
\end{abstract}
\maketitle



Electrons on the surface of a three-dimensional topological insulator (TI) are massless Dirac fermions with linear energy-momentum dispersion \cite{MooreReview, HasanReview, QiReview}. These electronic systems are promising for novel applications ranging from spin-based field effect transistors \cite{TIFET}, ultrafast opto-spintronic devices \cite{McIver} and high-speed quantum information processors \cite{FuKane}, whose performance depends critically on the dynamics of hot carriers. Unlike the case in graphene \cite{Graphene}, helical Dirac fermions in a TI interact not only with phonons but also with an underlying bulk reservoir of electrons \cite{HasanReview}. Therefore it is important to understand separately their coupling mechanisms to each of these degrees of freedom. However, high frequency optical conductivity \cite{HancockAguilar} and ultrafast optical experiments \cite{Kumar,HsiehSHG} do not directly separate bulk and surface signals nor do they distinguish different relaxation channels.


Time- and angle-resolved photoemission spectroscopy (TrARPES) is a powerful technique to study carrier dynamics with energy and momentum resolution. A recent TrARPES study has shown that persistent surface carrier population in a TI can be induced by photo-excitation \cite{Sobota}. In this letter, we focus on disentangling different relaxation channels following photo-excitation over a range of electron doping suitable for device applications. Analysis of the TrARPES spectra enables us to separately obtain temperature and chemical potential relaxation of both surface and bulk. Our data reveal that interband inelastic electron-electron (e-e) scattering is suppressed and surface-to-bulk coupling is mediated entirely by phonons. At low lattice temperature, this latter scattering channel turns off and we observe that the surface temperature decay rate scales as a power law with doping indicating acoustic phonon assisted cooling of 2D Dirac fermions.

\begin{figure*}[t]
\includegraphics[scale=0.84]{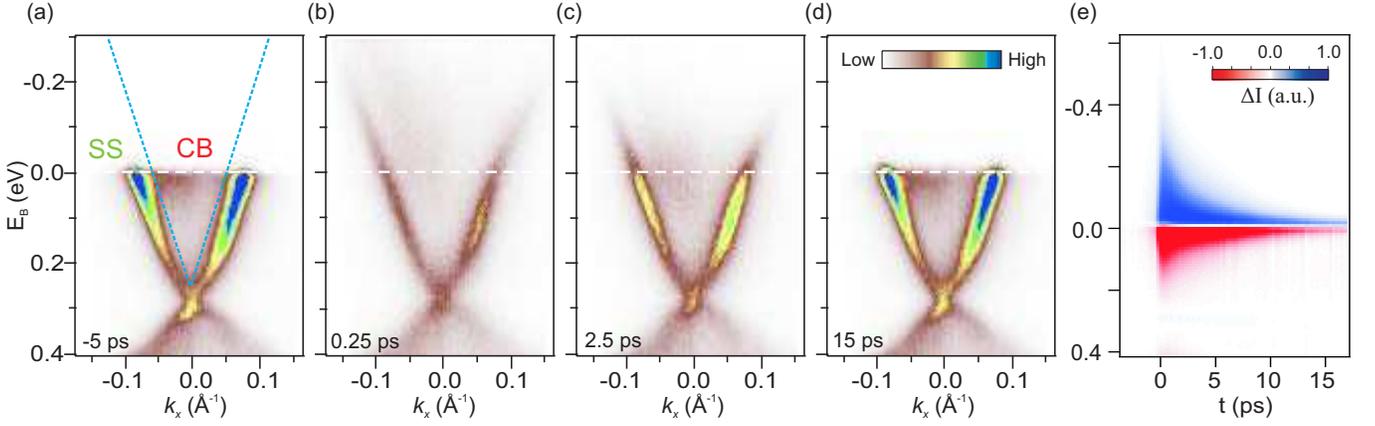}
\caption{TrARPES spectra of Bi$_2$Se$_3$. (a)-(d) Energy-momentum spectra sliced through
3D data volume along $\Gamma$-M direction at -5 ps (a), 0.25 ps (b), 2.5
ps (c) and 15 ps (d) taken at 15 K on an $E_D$ = 0.28 eV sample. (e) Momentum integrated difference spectra subtracting the spectrum at $t$ = -5 ps. The blue dashed lines in (a) define the boundary between conduction bulk band (CB) and surface state (SS) for the analysis performed in the text.}
\label{fig:Fig1}
\end{figure*}

Single crystals of Bi$_2$Se$_3$ are cleaved under ultra-high-vacuum ($<6\times10^{-11}$ Torr) \cite{SOM}. Their doping levels are determined by measuring their static ARPES spectra \cite{SOM}. TrARPES is performed following a pump-probe scheme \cite{Perfetti07} that uses 1.55 eV pulses with a fluence of 100 $\mu$J/cm$^2$ as a pump and 6.2 eV ultrafast laser pulses as a probe. A time-of-flight electron energy analyzer \cite{TOFARPES} is used to simultaneously collect the energy-momentum spectra $I(E,k_x,k_y,t)$ of Bi$_2$Se$_3$ at variable pump-probe time delay $t$ without sample or detector rotation \cite{BSSpinPRL}.

To understand the relaxation dynamics of the states near the Fermi level $E_F$, we show low energy ARPES spectra along the $\Gamma-M$ direction of the surface Brillouin zone at several time delays [Figs. 1(a)-(d)]. Fig. 1(a) shows the surface states (SS) and conduction bulk (CB) band of a Bi$_2$Se$_3$ sample with the Dirac point $E_D =$ 0.28 eV below $E_F$ before the pump pulse arrives ($t =$ -5 ps). The measured bandstructure is consistent with prior studies \cite{Xia}. The ARPES spectra taken immediately after the pump excitation [Fig. 1(b)] resemble a thermalized electron distribution at elevated temperature for both SS and CB. This hot distribution cools down progressively at longer time delays [Figs. 1(c) and (d)]. However, it can be seen from the difference of momentum integrated spectra $\Delta I(E,t) = I(E,t)-I(E,t=-5 ps)$ that the equilibrium has not been reached within the measurement time window [Fig. 1(e)].

\begin{figure}[t]
\includegraphics[scale=0.85]{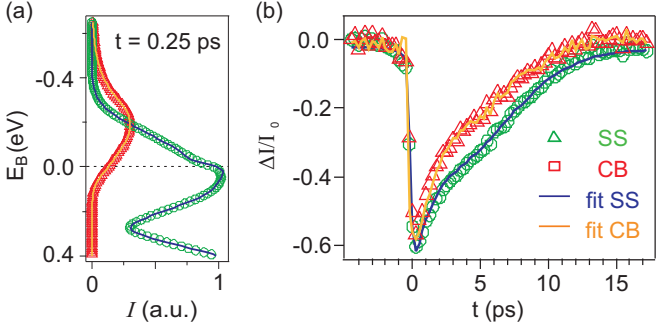}
\caption{Fitting of momentum-integrated spectra (a) Momentum integrated spectra for SS (green) and CB (red) and their respective fits (purple and orange lines) at $t =$ 0.25 ps for the data shown in Fig. 1. (b) Constant energy slices of the difference spectra as a function of $t$ at 0.1 eV for the same data and fit. The spectra are normalized over the pre-pump intensity $I_0 = I(t = -5 $ ps$)$.}\label{fig:Fig1fit}
\end{figure}

In order to quantitatively describe the relaxation dynamics of the hot bulk and surface populations in Fig. 1, we separately analyze the spectral intensities for the surface and the bulk. Fig. \ref{fig:Fig1fit}(a) shows $I(E,t)$ for SS and CB at $t =$ 0.25 ps obtained in their respective energy-momentum region [Fig. 1(a)]. We fit $I(E,t)$ for both SS and CB with the following equation \cite{IshidaFann} to extract their respective electronic temperatures $T_e(t)$ and chemical potentials $\mu(t)$ as a function of time:
\begin{eqnarray}
I(E,t) = A(t)\int_{-\infty}^{\infty}&d&\epsilon[f_{\textsc{fd}}(E,T_e(t),{\mu}(t))\times D(E)]\nonumber\\
    &G&(E-\epsilon,w(t))
\end{eqnarray}
Here $A(t)$ is a scaling factor \cite{IshidaFann}, $f_{\textsc{fd}}$ is the Fermi-Dirac distribution, $G(E,w)$ is a Gaussian function with energy independent width $w$ that includes the instrumental resolution and time-dependent broadening due to increased scattering rate after photo-excitation \cite{Fauster}. The time independent $D$, which is a product of density of states (DOS) and photoemission matrix element \cite{Hufner}, is obtained through a global best fit for all $t$ \cite{SOM}. Both the SS and CB fittings agree very well with data for all $t$ [Fig. \ref{fig:Fig1fit}(b)] (also see \cite{SOM}). The fact that both SS and CB populations can be described by Fermi-Dirac distributions at $t = $0.25 ps [Fig. \ref{fig:Fig1fit}(a)] suggests that intraband thermalization of SS and CB electrons is established within the instrumental resolution time ($\sim$ 200 fs).

\begin{figure*}[t]
\includegraphics[scale=1.1]{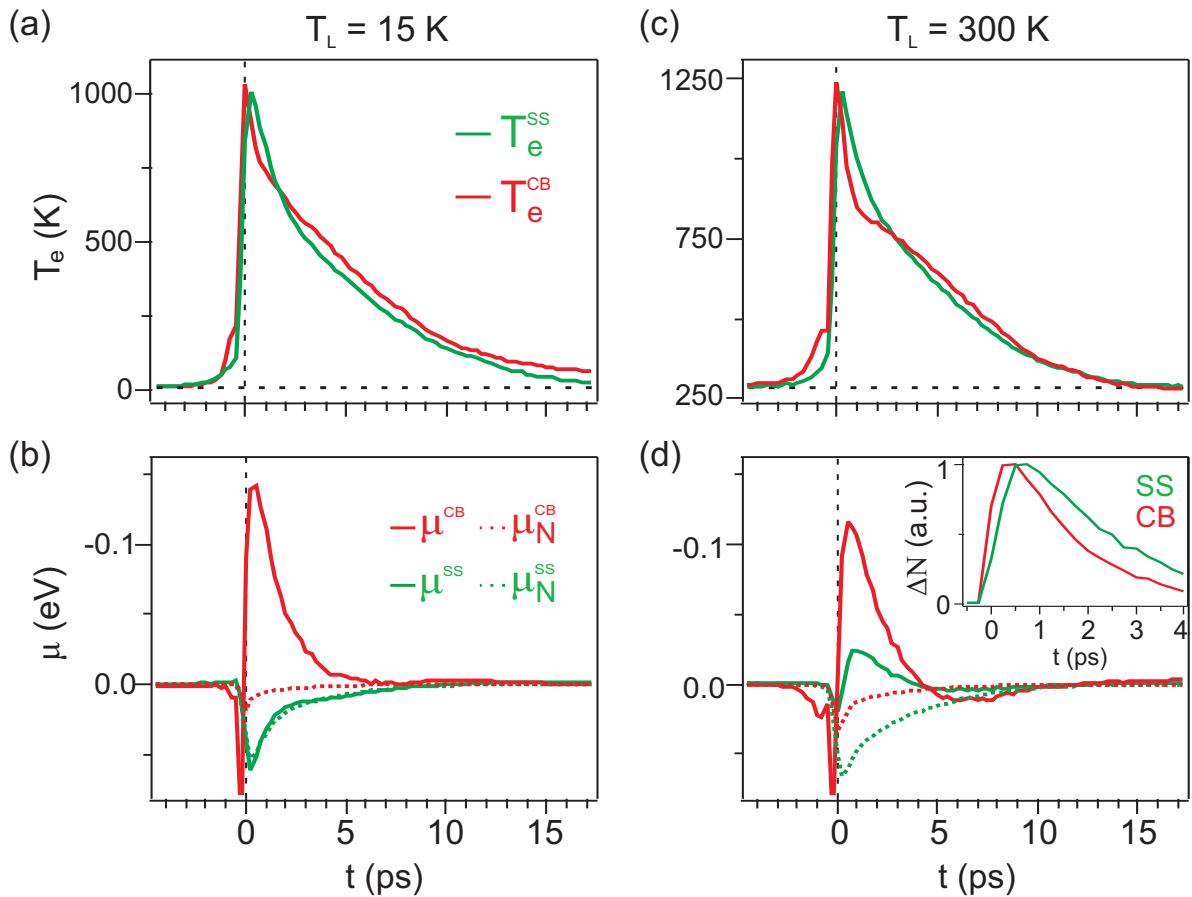}
\caption{Lattice temperature dependent interband surface-bulk electron dynamics. (a) Electronic temperature $T_e$ and (b) chemical potential $\mu$ for SS (green) and CB (red) at lattice temperature $T_L =$ 15 K. (c) and (d) are at $T_L =$ 300 K. The green dashed traces ${\mu}^{\textsc{ss}}_N$ are the SS chemical potential as a function of $T^{\textsc{ss}}_e$ under constant particle number. The red dashed traces are ${\mu}^{\textsc{cb}}_N$ for CB. The change of particle number is shown in (d) inset. (The features before $t = 0$ are due to the processes where the roles of pump and probe are reversed, which is commonly seen in TrARPES \cite{Perfetti07}.)} \label{fig:Fig2}
\end{figure*}

Fig. \ref{fig:Fig2} shows the obtained electronic temperatures and chemical potentials for both the SS and CB. We focus on the temperature first. As shown in Figs. \ref{fig:Fig2}(a), SS (CB) temperature $T^{\textsc{ss}}_e$ ($T^{\textsc{cb}}_e$) starts at the lattice temperature $T_L =$ 15 K at $t =$ -5 ps.  The maximum temperature increase of $\sim$ 1100K at $t =$ 0 ps allows us to estimate an electron specific-heat coefficient $\gamma$ of 2 mJ$/mol\cdot K^2$ (assuming 50 nm penetration depth of the pump pulse \cite{Sobota}). The relaxation time agrees with previous optical studies \cite{HsiehSHG}.

The analysis of the time dependence of the chemical potential reveals carrier population dynamics both within and between the bands. In general, chemical potential is a function of both temperature and particle number $N$ \cite{Kittel}. In the case where $N$ is fixed, we derive \cite{SOM} a relationship between chemical potential and our measured temperature assuming linear and parabolic dispersion for the SS and CB bands \cite{Kittel} respectively. The SS and CB chemical potentials derived this way (${\mu}^{\textsc{ss}}_N$ and ${\mu}^{\textsc{cb}}_N$) are plotted in Fig. \ref{fig:Fig2}(b) along with the chemical potentials directly obtained by fitting our data to Eq. (1) (${\mu}^{\textsc{ss}}$ and ${\mu}^{\textsc{cb}}$). Remarkably, a good overlap between ${\mu}^{\textsc{ss}}$ and ${\mu}^{\textsc{ss}}_N$ as a function of $t$ is observed at $T_L =$ 15 K [Figs. \ref{fig:Fig2}(b) green] indicating constant SS particle number $N^{\textsc{ss}}$. ${\mu}^{\textsc{cb}}$ is significantly higher than ${\mu}^{\textsc{cb}}_N$ [Figs. \ref{fig:Fig2}(b) red], suggesting instantaneous $N^{\textsc{cb}}$ increase upon photo-excitation. The increasing $N^{\textsc{cb}}$ shows that photo-excitation from deeper valence bands primarily populates CB, consistent with a related study that shows bulk dominant direct optical transition \cite{Sobota}. The fact that $N^{\textsc{ss}}$ stays constant indicates that there is no net particle transfer from CB to SS.

In order to understand whether the transfer of carriers between surface and bulk can happen at higher lattice temperature, we now investigate the chemical potentials and temperature dynamics at 300 K [Figs. \ref{fig:Fig2}(c) and (d)]. The most striking feature is that ${\mu}^{\textsc{ss}}$ is much bigger than ${\mu}^{\textsc{ss}}_N$ [Fig. \ref{fig:Fig2}(d)], indicating a strong increase of surface carrier density in contrast to 15 K [Fig. \ref{fig:Fig2}(b)]. To find where the extra surface carriers come from, we notice ${\mu}^{\textsc{cb}}$ is lower at 300 K than at 15 K [Fig. \ref{fig:Fig2}(b) and (d) solid red traces]. By calculating the change of particle numbers after photo-excitation of SS and CB ($\Delta N^{\textsc{ss}}$ and $\Delta N^{\textsc{cb}}$) \cite{SOM}, we see that the increase of $\Delta N^{\textsc{ss}}$ at 300 K has a slower rise time than $\Delta N^{\textsc{cb}}$ at 15 K [Fig. \ref{fig:Fig2}(d) inset]. This shows that SS particle number increase is not due to direct optical transition. Rather it is a result of transfer from CB to SS through phonon scattering at elevated lattice temperature due to higher phonon-scattering rate above Debye temperature ($\Theta_D$ = 182 K \cite{ShoemakeMills}) \cite{Grimvall}. The evidence for such phonon-assisted CB-SS scattering can also be seen in the temperature dynamics [Figs. \ref{fig:Fig2}(a) and (c)]. $T^{\textsc{ss}}_e$ and $T^{\textsc{cb}}_e$ only equilibrate at 300 K within the probed time window, indicating that phonons scatter carriers, transferring energy and mediating the thermalization between SS and CB at 300 K. Since inelastic e-e scattering is mainly responsible for ultrafast thermalization \cite{Grimvall} within a single band, the lack of thermalization between SS and CB at 15 K  [Figs. \ref{fig:Fig2}(a)] suggests that inelastic e-e scattering time between SS and CB exceeds the measured time window. (See \cite{SOM} for more $T^{\textsc{ss}}_e$ vs. $T^{\textsc{cb}}_e$ at 15 K.) Such suppression of interband inelastic e-e scattering is likely due to the kinematics constraints \cite{Landsberg}.

\begin{figure}[t]
\includegraphics[scale=0.68]{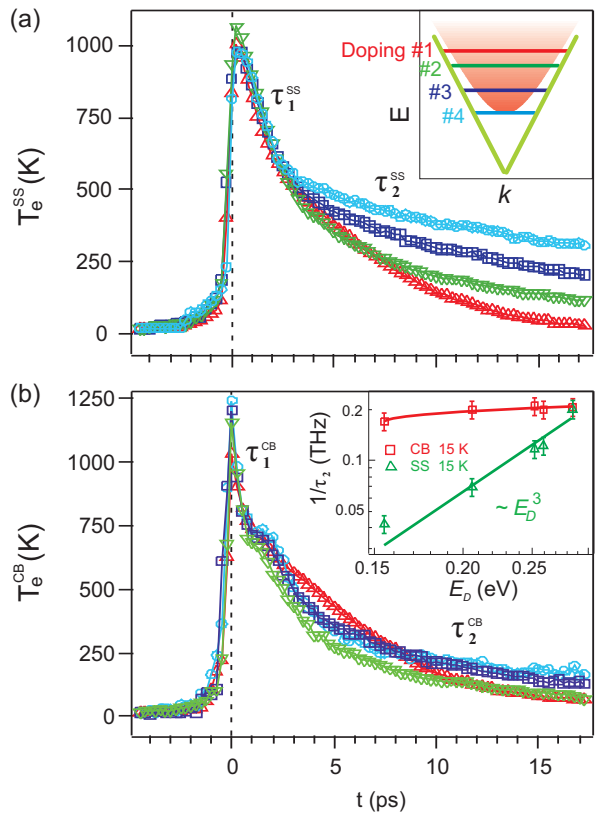}
\caption{Electron density dependent intraband cooling of surface Dirac fermions. (a) Electronic temperature as a function of $t$ of surface electrons with different $E_D$ at 15 K. Different colors corresponds to $E_D$ in matching color in the inset. (b) $T^{\textsc{cb}}$ at 15 K. Inset: Inverse cooling time of the slow component of $T_e$ as a function of $E_D$ at 15 K for SS (green triangles) and CB (red squares). $1/{{\tau}^{\textsc{ss}}_2} \propto {E_D}^3$ at 15 K.} \label{fig:Fig3}
\end{figure}

Having shown that interactions between SS and CB are suppressed at 15 K [Figs. \ref{fig:Fig2}(a) and (b)], we proceed to investigate their respective intraband cooling mechanisms. Because different mechanisms can be distinguished based on carrier density dependence, as demonstrated in other Dirac materials \cite{Bistritzer}, we measure the intraband cooling dynamics of Bi$_2$Se$_3$ over a wide range of dopings \cite{SOM}. To vary the doping, we used samples from two different batches with high and low carrier concentrations along with surface doping effect \cite{King,ParkHsieh}. The surface doping effect has been commonly observed in topological insulators such as Bi$_2$Se$_3$ \cite{King} and is attributed to either adsorption or moving of Se vacancies towards the surface \cite{ParkHsieh}. Figure \ref{fig:Fig3}(a) shows that the surface temperature has two decay components $\tau^{\textsc{ss}}_1$ and $\tau^{\textsc{ss}}_2$. The fast-decaying component $\tau^{\textsc{ss}}_1$ is independent of $E_D$. This is consistent with optical phonon cooling, which is insensitive to doping as observed in graphene \cite{Huang}. When the electronic temperature cools below $\sim 600 K$, the optical phonon cooling becomes less effective. Instead, the decay becomes dominated by a density dependent component $\tau^{\textsc{ss}}_2$ whose rate increases with electron doping [Fig. \ref{fig:Fig3}(a)]. In contrast, Figure \ref{fig:Fig3}(b) shows that bulk temperature dynamics is qualitatively different. The fast component $\tau^{\textsc{cb}}_1 =$ 0.7$\pm$0.1 ps [Fig. \ref{fig:Fig3}(b)] matches with a previous ultrafast study, which has been attributed to an optical phonon mediated intraband cooling of CB \cite{HsiehSHG}. The slow component $\tau^{\textsc{cb}}_2$ exhibits no discernable dependence on $E_D$ [Fig. \ref{fig:Fig3}(b)].

We use the density dependence of the slow component to reveal its microscopic origin. We extract $\tau^{\textsc{ss}}_2$ through a biexponential fit of $T^{\textsc{ss}}_e(t)$ \cite{SOM} and plot it as a function of doping $E_D$ in Fig. \ref{fig:Fig3}(b) inset. Such exponential decay of the electronic temperature in a Dirac system has been previously predicted for graphene \cite{Bistritzer} and also observed in graphite \cite{Kampfrath} and recently Bi$_2$Se$_3$ \cite{Sobota}. The SS cooling rate scales as $1/{\tau}^{\textsc{ss}}_2 \propto {E_D}^{3 \pm0.5}$, which is in good agreement with the theoretical calculations of cooling of 2D Dirac fermions via acoustic phonons \cite{Bistritzer}. In stark contrast, $1/\tau^{\textsc{cb}}_2$ exhibits little dependence on doping, which agrees with theoretical predictions that acoustic phonon-mediated carrier cooling in 3D metals is density independent \cite{Allen}. Consistent with the phonon assisted surface bulk coupling presented earlier, the cooling rate at 300 K lies in between that of the SS and CB at 15 K \cite{SOM}. It is worth mentioning that in the low doping regime where $\mu << k_{B}T_e$, acoustic phonon cooling of Dirac fermions will be further suppressed and is predicted to have a different decay dynamics \cite{Bistritzer}. As a result, the doping dependence of the slow decay component may deviate from the power at low doping levels.

In conclusion, we have directly visualized different scattering channels of surface and bulk electrons on a TI with TrARPES. At high temperature, phonons are largely responsible for scattering electrons between the surface states and
the conduction bulk band. At low temperature, the coupling between surface and bulk is suppressed. We further reveal the surface cooling rate follows a power law dependence on the carrier density, which is a signature of acoustic phonon mediated cooling of 2D Dirac fermions. This suggests that by tuning to the charge neutrality point one can enter a regime where acoustic phonon scattering of Dirac fermions is completely eliminated and transport properties are determined solely by structural and chemical defects \cite{Steinberg,SeoBeidenkopf}. An exciting possibility is the creation of long-lived hot photo-carriers that can be used for high-efficiency photothermoelectric applications \cite{McIver,Gabor}.

The authors would like to thank Nathan Gabor and Justin Song for useful discussions and James McIver, Alex Frenzel and Fahad Mahmood for careful reading of the manuscript. This research is supported by Department of Energy Office of Basic Energy Sciences Grant numbers DE-FG02-08ER46521 and DE-SC0006423 (data acquisition and analysis), Army Research Office (ARO-DURIP) Grant number W911NF-09-1-0170 (ARTOF spectrometer) and in part by the MRSEC Program of the National Science Foundation under Grant number DMR - 0819762 (initial feasibility study). P.J-H. acknowledges support from the U.S. Department of Energy, Office of Basic
Energy Sciences, Division of Materials Sciences and Engineering, Early Career Award number DE.SC0006418 and a Packard Fellowship (material growth and characterization).


\end{document}